\documentclass[twocolumn]{aastex631}

\emergencystretch=\maxdimen
\begin{document}

\title{Massive Double White Dwarf Binary Mergers from the Moon: 
Extending the Reach of Multi-messenger Astrophysics}




\newcommand{\note}[1]{{\textcolor{red}{\sf{[#1]}}}}

\correspondingauthor{Manuel Pichardo Marcano}
\email{mmarcano@fisk.edu}

\author[0000-0003-4436-831X]{Manuel Pichardo Marcano}
\affiliation{Department of Life and Physical Sciences, Fisk University, 1000 17th Avenue N., Nashville, TN 37208, USA}
\affiliation{Vanderbilt Lunar Labs Initiative, Vanderbilt University, 6301 Stevenson Center Lane, Nashville, TN 37235, USA}

\author[0000-0002-8065-1174]{Anjali B. Yelikar}
\affiliation{Vanderbilt Lunar Labs Initiative, Vanderbilt University, 6301 Stevenson Center Lane, Nashville, TN 37235, USA}

\author[0000-0003-1007-8912]{Karan Jani}
\affiliation{Vanderbilt Lunar Labs Initiative, Vanderbilt University, 6301 Stevenson Center Lane, Nashville, TN 37235, USA}



\begin{abstract}
We explore the potential of lunar-based gravitational-wave detectors to broaden the multi-messenger astrophysics landscape by detecting mergers of massive ($M_1,M_2 >1 M_\odot$) double white dwarf (WD) binaries. These systems are potential progenitors of Type Ia supernovae and could serve as independent probes of cosmic expansion. We examine two proposed lunar gravitational-wave detector concepts operating in the sub-hertz band (0.1–1 Hz): the Gravitational-Wave Lunar Observatory for Cosmology (a proxy for suspended test mass detectors) and the Lunar Gravitational-Wave Antenna (a proxy for seismic array detectors). We estimate that these detectors could reach distances of up to $\sim$1 Gpc for the most massive mergers. We show that lunar detectors could observe up to dozens of massive WD mergers annually, including those originating from globular clusters. Lunar detectors would constrain the masses of these WDs with an unprecedented accuracy of one part in a million. Furthermore, these detectors would provide early warnings of weeks before merger, including sky-localization of square arcminute resolution, enabling a new era of coordinated multi-messenger follow-up of electromagnetic transients—whether they evolve into Type Ia supernovae or accretion-induced collapse events.

\end{abstract}

\keywords{}


\section{Introduction} \label{sec:intro}

Type Ia supernovae play a crucial role in modern cosmology, serving as standard candles for measuring cosmic distances and probing the expansion history of the Universe \citep[for a review see][]{Ruiter(2025)A&ARv2025A&ARv..33....1R}. However, the precision of Type Ia supernovae as distance indicators is currently limited by systematic uncertainties, including the unknown nature of their progenitor systems \citep[e.g.][]{Kirshner(2010)dken2010dken.book..151K,Vincenzi(2024)ApJ2024ApJ...975...86V}. Resolving the progenitor problem is not only critical for improving the accuracy of Type Ia supernovae as cosmological probes but also for understanding the observed discrepancies between local and early Universe measurements of the Hubble constant, known as the Hubble tension \citep[see][and references therein]{BousisHubbleTension}.

Two main progenitor scenarios have been proposed for Type Ia supernovae: the single-degenerate (SD) scenario, involving a white dwarf (WD) accreting material from a non-degenerate companion \citep{vandenHeuvel1992A&A...262...97V,Rappaport1994ApJ...426..692R,Li1997A&A...322L...9L,Langer2000A&A...362.1046L,Han2004MNRAS.350.1301H}, and the double-degenerate (DD) scenario, involving the merger of two WDs \citep{Iben1984ApJS...54..335I,Webbink1987fbs..conf..445W} In this scenario, the merger of two sufficiently massive WDs can lead to a Type Ia supernovae if the combined mass exceeds the Chandrasekhar mass ($1.4~M_\odot$). Sub-Chandrasekhar DD mergers have also been proposed as possible progenitors for Type Ia supernovae \citep{Woosley(1980)tsup1980tsup.work...96W,Nomoto(1982)ApJ1982ApJ...257..780N} through the double-detonation mechanism, where ignition of the helium layer triggers secondary carbon detonation in the core \citep[e.g.][]{vanKerkwijk(2010)ApJL2010ApJ...722L.157V,Ruiter(2011)MNRAS2011MNRAS.417..408R,2014ApJ...797...46S}.



A Type Ia supernova is not the only outcome of the merger of two WDs. Other merger products include massive WDs \citep{Dunlap(2015)ASPC2015ASPC..493..547D,Ferrario(1997)MNRAS1997MNRAS.292..205F,Kawka(2023)MNRAS2023MNRAS.520.6299K} and other peculiar WDs \citep{Raddi(2019)MNRAS2019MNRAS.489.1489R,Vennes(2017)Sci2017Sci...357..680V,Shen(2018)ApJ2018ApJ...865...15S}. Massive oxygen-neon-rich (ONe) WDs could also collapse to form a neutron star (NS) through accretion-induced collapse (AIC) \citep[e.g.][]{Nomoto(1979)wdvd,Miyaji(1980)PASJ}. The collapse can be due to  accretion of hydrogen-rich material \citep{Hurley(2010)MNRAS2010MNRAS.402.1437H} or from a DD merger \citep{Saio(1985)A&A1985A&A...150L..21S,Ivanova(2008)MNRAS2008MNRAS.386..553I,Wu(2018)RAA2018RAA....18...36W}. Although no direct detection of an AIC event has been made to date, this scenario has been invoked to explain several observed phenomena, such as the large fraction of NS in globular clusters (GCs) and the formation of recycled pulsars with low space velocities \citep[e.g.][]{Bailyn1990PulsarsfromAIC,Kitaura2006,Dessart2006}. For a review on the  formation of NS systems through AIC in WD binaries  see \citet{WangAIC2020Review}. Moreover, AIC is expected to produce a variety of electromagnetic transients all the way from Radio \citep[e.g.][]{Piro2013ApJ...762L..17P} to Gamma Rays \citep[e.g.][]{Lyutikov(2017)GammaRay}. 


\begin{figure}[t!]
\includegraphics[width=1.05\columnwidth]{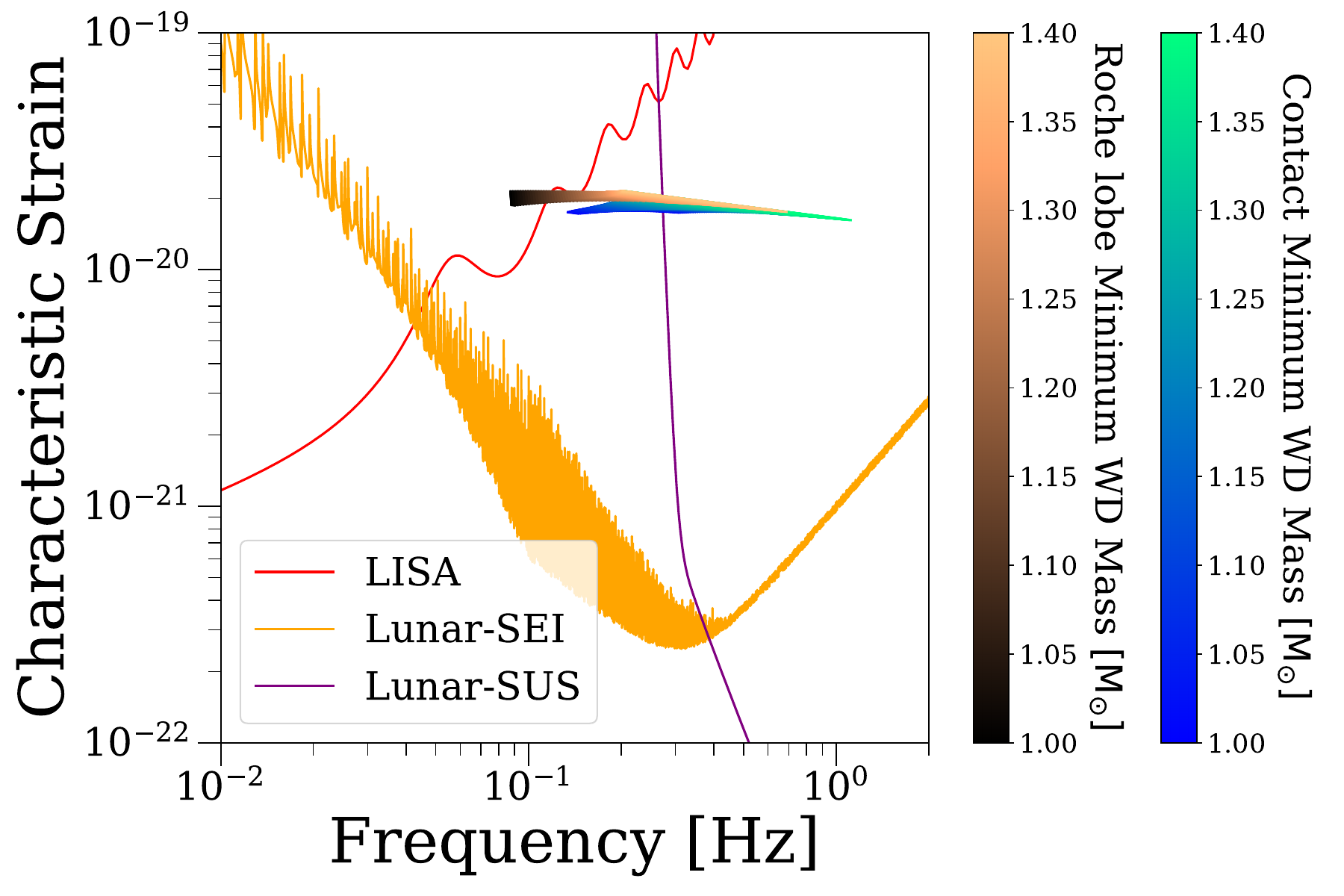}
\caption{Evolution of the characteristic strain as a function of gravitational-wave frequency over one year before merger for all massive (individual masses $M>1 M\odot$) WD mass combinations at a distance of 10 Mpc. The color gradient represents the lower component mass for each  double WD binary considered. We consider two distinct merger frequency definitions: the Roche lobe case in brown and the contact case in green/blue (see sec~\ref{sec:method}). The sensitivity curves of \texttt{Lunar-SUS}, LISA, and \texttt{Lunar-SEI} are overlaid in purple, red, and orange, respectively. } 
\label{fig:PlotStrain10MpcBothCasesinOne}
\end{figure}

Either outcome of the merger of massive WDs, Type Ia supernova or the formation of a NS via AIC,  are expected to produce distinctive electromagnetic signatures making these events ideal targets for coordinated gravitational-wave and electromagnetic observations. The future space-borne Laser Interferometer Space Antenna \citep[LISA;][]{ColpiLisa2024} will be capable of detecting gravitational waves within a frequency range of 0.1 mHz to 1 Hz. However, LISA's sensitivity decreases at higher frequencies \citep{RobsonCOrnishLISASen}, where we anticipate the merger of most double WD binaries \citep{Maselli(2020)WDDecihertz,Zou(2020)RAA2020RAA....20..137Z} (with $f_{merge} \gtrsim 0.1$ Hz).

Despite this limitation, LISA is expected to offer significant insights into Type Ia supernovae from progenitor systems within our Milky Way and the Magellanic Clouds \citep{Kopparapu(2007)ApJ,KorolLISASN24,Criswell(2025)PhRvD}. LISA may also enable indirect detection of mergers by observing the disappearance of nearly monochromatic gravitational waves associated with these events \citep{SetoLISA2023}. LISA’s primary limitation is that it can only study Galactic and nearby (Magellanic clouds) double WD mergers so to explore larger volumes and capture events in other galaxies, we need detectors with enhanced sensitivity, especially near sub-hertz frequencies. Previous studies of WD binaries in the sub-hertz frequencies include \citet{Mandel2018CQGra..35e4004M,Huang2020PhRvD.102f3021H,Sedda(2020)Decihertz,Maselli(2020)WDDecihertz,Yoshida2021,Izumi(2021)hgwa}, but these studies have focused mainly on space-based detectors like DECIGO \citep{DecigoState2021PTEP.2021eA105K}, TianQin \citep{TianQin2016CQGra..33c5010L} and LISA \citep{ColpiLisa2024} and also on Type Ia supernova progenitors, without special attention to the merger of the most massive WDs that could give rise to AIC events.

The Moon's low seismic noise levels \citep{Lognonne(2009)JGRE2009JGRE..11412003L,Majstorovic(2025)PhRvD2025PhRvD.111d4061M} makes lunar-based gravitational-wave detectors ideal for accessing the sub-Hertz frequency band (0.1-1 Hz), where WD binaries merge and LISA's sensitivity decreases (see Figure~\ref{fig:PlotStrain10MpcBothCasesinOne}). Detectors utilizing either suspended test masses or seismometer arrays have been proposed as future gravitational-wave observatories \citep{Amaro-Seoane(2021)CQGra2021CQGra..38l5008A,GlocJani2020arXiv200708550J,LGWA(2021),Li(2023)SCPMA2023SCPMA..6609513L}. The scientific potential of the sub-hertz band extends beyond WD mergers, with recent work demonstrating detection prospects for core-collapse supernovae signals in this regime \citep{Gill(2024)arXiv}. For a recent review on lunar gravitational-wave detection see \citet{Branchesi2023SSRv..219...67B} and \citet{Cozzumbo(2024)RSPTA}.


The Lunar Gravitational-wave Antenna \citep[LGWA;][]{LGWA(2021),LGWAWP} is a proposed future gravitational-wave detector on the Moon. Given the size of the Moon and the expected noise produced by the lunar seismic background, the LGWA would be able to observe GWs from about 0.001 Hz to 1 Hz (where LISA's sensitivity decreases, see Figure~\ref{fig:PlotStrain10MpcBothCasesinOne}). The mission consists of deploying an array of high-end seismometers (SEI) on the Moon to monitor normal modes of the Moon. \textbf{We use LGWA as a proxy for measuring gravitational waves through lunar normal modes, referred in this study as \texttt{Lunar-SEI}. }



The Gravitational-wave Lunar Observatory for Cosmology \citep[GLOC;][]{GlocJani2020arXiv200708550J, Jani_Artemis2022, ballmer2022snowmass2021cosmicfrontierwhite}, is a proposed detector with suspended test masses and interferometeric readout to detect gravitational waves  in the sub-Hertz frequency range (where LISA's sensitivity decreases, see Figure~\ref{fig:PlotStrain10MpcBothCasesinOne}). The detection principles are similar to that of current terrestrial gravitational-wave detectors Advanced LIGO \citep{LIGOScientific:2014pky}, Advanced Virgo \citep{VIRGO:2014yos} and KAGRA \citep{KAGRA:2020tym}, but without requiring large-scale infrastructure. The setup will consist of three end-stations around a lunar crater. \textbf{We use GLOC as a proxy for measuring gravitational waves through suspended test mass (SUS), referred in this study as \texttt{Lunar-SUS}.}


The Laser Interferometer Lunar Antenna (LILA) project is a future detector concept that aims to measure gravitational waves from deci-hertz to several hundreds of hertz using both the suspended test mass \texttt{Lunar-SUS} and strain-meter measurement of lunar modes \citep[\texttt{Lunar-SEI};][]{Jani(2024)AAS,  Logn(2024)LPICo}. The LILA project could be deployed on the lunar surface within the 2030s by the NASA Commercial Lunar Payload Services and Artemis programs~\citep{Trippe(2024)LPICo}.

In this work, we focus on the mergers of massive WDs ($M_{1},M_2 > 1~M_\odot$) and explore the capabilities of future lunar-based gravitational-wave observatories. 
In Section~\ref{sec:method} we describe our methods to find the merger frequencies of double WD binaries, their sky-averaged detection distances and estimation on their intrinsic (masses) and extrinsic (sky-location, distance) parameters. In Sections~\ref{sec:results} and \ref{sec:discussion}, we present our findings on detection capabilities and early warning potential, demonstrating that lunar-based gravitational-wave detectors could observe dozens to thousands of merger events annually. We discuss the implications for multi-messenger astronomy, including electromagnetic follow-up opportunities and new constraints on both Type Ia supernova progenitors and AIC events.








\section{Method}
\label{sec:method}

We limit our study to the merger of massive WDs and define them as those with masses $M > 1~M_\odot$. We consider only those binaries were $M_{total} = M_1 + M_2 \ge 2~M_\odot$  with component masses in the range $1~M_\odot$ to $1.4~M_\odot$. We evolve these binaries for one year before the merger. Extended mission lifetimes of 5-10 years would increase the detection distance by a factor of approximately $1.5\times$ depending on the masses of the WDs for \texttt{Lunar-SEI} by allowing it to track inspiraling binaries from lower initial frequencies ($f<0.1$ Hz). However we adopt a one year limit to focus on multi-messenger prospects, particularly in Type Ia supernova and AIC electromagnetic counterpart search and follow-up. We define the merger in two ways, the first one is what we call the ``contact case" where we define the merger frequency, $f_{max}$, to be when the system orbital separation is equal to the sum of the two WDs radii. This merger condition is not physically realistic, as WDs would undergo significant tidal disruption before reaching this separation. However, we include it as a theoretical upper limit and to maintain consistency with previous studies of DD Type Ia supernova progenitors that used this criterion as the merger condition \citep[e.g.][]{Maselli(2020)WDDecihertz,Li(2023)SCPMA2023SCPMA..6609513L,KinugawaDecihertztypeIa2022ApJ...938...52K}. For the second case, or ``Roche lobe case", the merger frequency is obtained by setting the Roche lobe radius equal to the radius of the less massive WD. For both cases, we approximate the gravitational wave from a detached double WD binary as nearly monochromatic emissions from two point masses, $M_1$ and $M_2$, in a circular orbit. For a circular orbit, the gravitational-wave frequency, $f_{gw}$, is twice the orbital frequency, $f_{orb}$, such that $f_{gw} = 2 \times f_{orb} = 2/P_{orb}$. Using Kepler’s law, we calculate the gravitational-wave frequency for a given orbital period, $P_{orb}$:
\begin{equation}
f_{gw} = \frac{1}{\pi} \times G^{1/2} \times (M_1 + M_2) ^{1/2} \times a^{-3/2}
\label{eq:fgw}
\end{equation}
where $a$ is the semi-major axis. 

The radius of a WD, $R_{wd}$, is calculated using Eggleton’s zero-temperature mass-radius relation as given by \citet{Verbunt1988}:

\begin{eqnarray}
\frac{R_{wd}}{R_\odot} = 0.0114 \left [  \left ( \frac{M}{M_{ch}} \right ) ^{-2/3} - \left ( \frac{M}{M_{ch}} \right ) ^{2/3} \right ] ^{1/2} \nonumber \\
    \times \left [ 1+ 3.5 \left ( \frac{M}{M_p} \right ) ^{-2/3} + \left ( \frac{M}{M_p} \right ) ^{-1}  \right ] ^{-2/3} ,
    \label{eq:wdmass}
\end{eqnarray}

where $M_{ch} = 1.44~M_\odot$ and $M_p = 0.00057~M_\odot$.

For the contact case we set $a =R_{WD}(M_1) +R_{WD}(M_2)$ in  Equation~\ref{eq:fgw} and use  Equation~\ref{eq:wdmass} for the radii. For the Roche lobe case, to determine the merger frequency, we compute the Roche lobe radius, $R_L$, using the formula from \citet{Paczyski1967}:

\begin{equation}
R_L = 2 \times 3^{-4/3} \times M_1^{1/3}\times (M_1+M_2)^{-1/3}\times a
 \label{eq:RL}
\end{equation}

where $M_1 < M_2$. By combining Equations~\ref{eq:RL} and \ref{eq:fgw}, we derive the gravitational-wave frequency at the merger, $f_{RL}$, as a function of the less massive WD, $M$:

\begin{equation}
    f_{RL} = \frac{2^{3/2} G^{1/2} M^{1/2}}{9 \pi R(M)^{3/2}}
    \label{eq:gwroche}
\end{equation}

where $R(M)$ is the WD radius calculated from Equation~\ref{eq:wdmass}.

Equation~\ref{eq:RL} is only an approximation, but this formula agrees with the tabulation of \cite{Kopal(1959)cbs1959cbs..book.....K} to within $\sim 3 \%$.


\subsection{Signal-to-Noise}

To evolve the combinations of double WD binaries until the merger we use the Python package LEGWORK~\citep{LEGWORK_apjs}\footnote{\url{https://legwork.readthedocs.io/en/latest/}}. Following \citet{PhysRevD.57.4535} and \citet{LEGWORK_apjs}, we calculate the position-, orientation-, and angle-averaged Signal-to-Noise ratio (SNR) for a circular binary system,  $\langle  \rho\rangle$, defined as: 

\begin{equation}
\langle  \rho\rangle^2=     \int^{f_{max}} _{f_{min}}  df  \frac {h_ {c}^ {2}}{f^ {2}S_ {n}(f)}  ,
\label{eq:snr}
\end{equation}

where $h_c$ is the characteristic strain amplitude, $S_n$ denotes the noise power spectral density of the detector, $f_{min}$ is the frequency one year before merger and $f_{max}$ is the merger frequency as defined in Equations~\ref{eq:fgw} and \ref{eq:gwroche} for the contact and Roche lobe cases respectively. We find $f_{min}$ by integrating the semi-major axis (a) of a circular binary as a function of time using Equation 5.9 from \citet{Peters1964} as implemented in LEGWORK~\citep{LEGWORK_apjs}. We calculate $h_c$ using the LEGWORK package for each mass combination. For the noise power spectral density of the \texttt{Lunar-SUS} detector, we use the optical noise curve of GLOC, which has a lower frequency limit of $f_{low} = 0.25$ Hz and it is reported in \citet{GlocJani2020arXiv200708550J}\footnote{\url{https://doi.org/10.5281/zenodo.3948466}}. For the \texttt{Lunar-SUS} detector we use the optimal design, referred to as the silicon model, of LGWA which was obtained from the GWFish repository~ \citep{GWFishPaper}\footnote{\url{https://github.com/janosch314/GWFish}}.

\subsection{Sky-Averaged Detection Distance}

We generated 100 WD masses evenly spaced from 1. to 1.4 $M_\odot$. We then computed the sky-averaged detection distance for all 4950 possible unique mass pairs ($M_1$, $M_2$). The sky-averaged detection distance represents the maximum distance at which the averaged SNR of a source (Equation~\ref{eq:snr}) is still above some detectable threshold. These are evolving sources where SNR accumulates as a function of time. For the average detection distance calculation, we integrate Equation~\ref{eq:snr} over a one year period, setting $f_{min}$ as the frequency one year before merger and $f_{max}$ to be the merger frequency. This distance is calculated by scaling the SNR at a given distance, $D$, to the detection threshold, $\rho_{th}$. The sky-averaged detection distance, $D_{avg}$, is defined as:
\begin{equation}
D_{avg} = \frac{\langle  \rho (D) \rangle}{\rho_{th}} \times D
\label{eq:HD}
\end{equation}
where $\langle  \rho (D) \rangle$ is the sky-average SNR at a luminosity distance of $D$, and $\rho_{th}$ is the SNR above which we consider a source detectable. For the rest of the paper and analysis, we fix $\rho_{th} = 8$.



\subsection{Parameter Estimation Setup}

To provide a forecast for the parameter estimation of future events, we use the Fisher information matrix formalism \citep{Finn1992PhRvD..46.5236F,Cutler1994PhRvD..49.2658C}. This method provides a lower bound on statistical measurement uncertainties in the high SNR limit. For our analysis, we use the Python package GWFish \citep{GWFishPaper}. To model the chirping gravitational wave, we use the IMRPhenomPv2 waveform~\citep{HannamWaveForm2014PhRvL.113o1101H}. Most of the work in the literature uses either a Newtonian approach to get the frequency evolution which will only model the inspiral regime well or use waveform models developed for binary black holes (BBH). The latter is what is used in this work and comes with its caveats that it ignores the finite size effects, but the inspiral of DWD can be captured by this model hence we use it for parameter estimation. The minimum frequency for likelihood integration is determined for the double WD system when they are one year away from merger. The maximum frequency considered is when the double WD merges, as defined in equations \ref{eq:fgw} and \ref{eq:gwroche} for the contact and Roche lobe cases respectively.

\begin{figure}[t!]
\includegraphics[width=1.0\columnwidth]{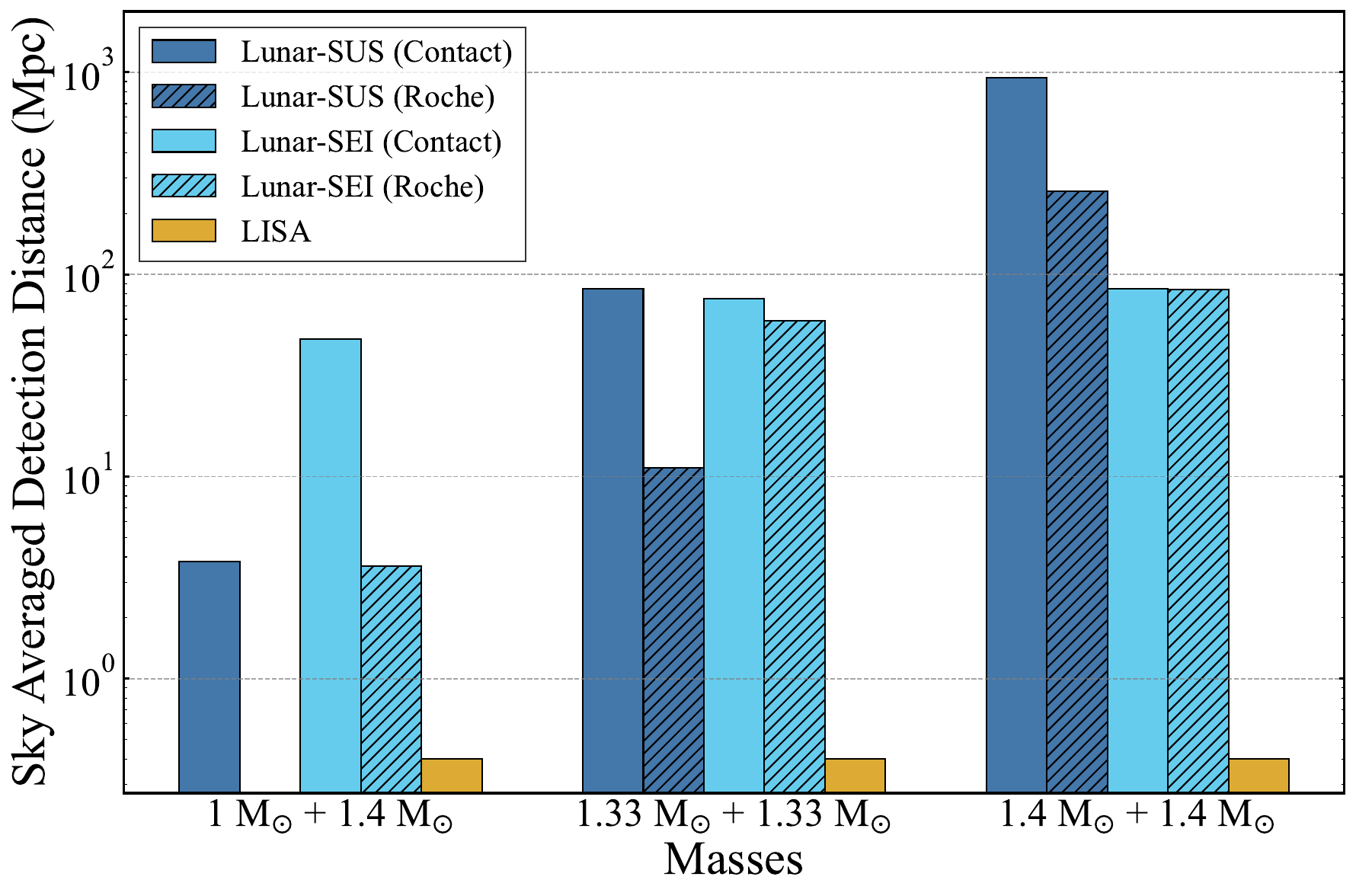}
\caption{One year sky-averaged detection distance for three different WD mass combinations. For the average detection distance calculation, we integrate Equation~5 over a one year period, setting $f_{min}$ as the frequency one year before merger and $f_{max}$ to be the merger frequency. We consider both the contact and Roche lobe cases (see sec~\ref{sec:method}) respectively. We reiterate that the contact case is not physically realistic, as WDs would undergo significant tidal disruption before merger and should be seen as an idealized upper limit for detection distance. Results are shown for both \texttt{Lunar-SUS} and \texttt{Lunar-SEI} detectors, comparing the contact case (solid bars) and Roche lobe overflow case (hatched bars) scenarios. LISA's detection range (orange) remains nearly constant at $\sim 0.4$ Mpc across all mass combinations.}  
\label{fig:HRDistBothCases}
\end{figure}

\section{Results}
\label{sec:results}

We present the results for both cases-the contact case and the Roche lobe case, for both the \texttt{Lunar-SUS} and \texttt{Lunar-SEI} detectors.

\subsection{Characteristic Strain}

In the Roche lobe case, we consider the merger to occur when the less massive WD fills its Roche lobe. For this case, the merger frequency only depends on the average density of the less massive (larger radius) WD and is given by Equation~\ref{eq:gwroche}. Figure~\ref{fig:PlotStrain10MpcBothCasesinOne} shows the one year strain evolution for all simulated WD merger combinations located at 10 Mpc for the Roche lobe case in brown and the color gradient represents the mass of the less massive  WD in each binary system. In the contact case scenario, we assume that the merger occurs when the separation between the two WDs is equal to the sum of their radii, i.e., $a = R_{wd}(M_1) + R_{wd}(M_2)$. We then evolve the system backward in time to determine the gravitational wave frequency one year before merger. The green/blue color gradient represents the mass of the less massive (larger radius) WD in each binary system for the contact case.  The sensitivity curves of \texttt{Lunar-SUS}, LISA, and \texttt{Lunar-SEI} are overlaid in purple, red, and orange, respectively. This figure demonstrates that \texttt{Lunar-SEI} and \texttt{Lunar-SUS} have the potential to detect the final stages of massive WD mergers at frequencies above 0.1 Hz, where LISA's sensitivity diminishes.


\subsection{Sky-Averaged Detection Distance}

\begin{figure*}
\centering
\includegraphics[trim = 0 30 0 50, clip, scale=0.5]{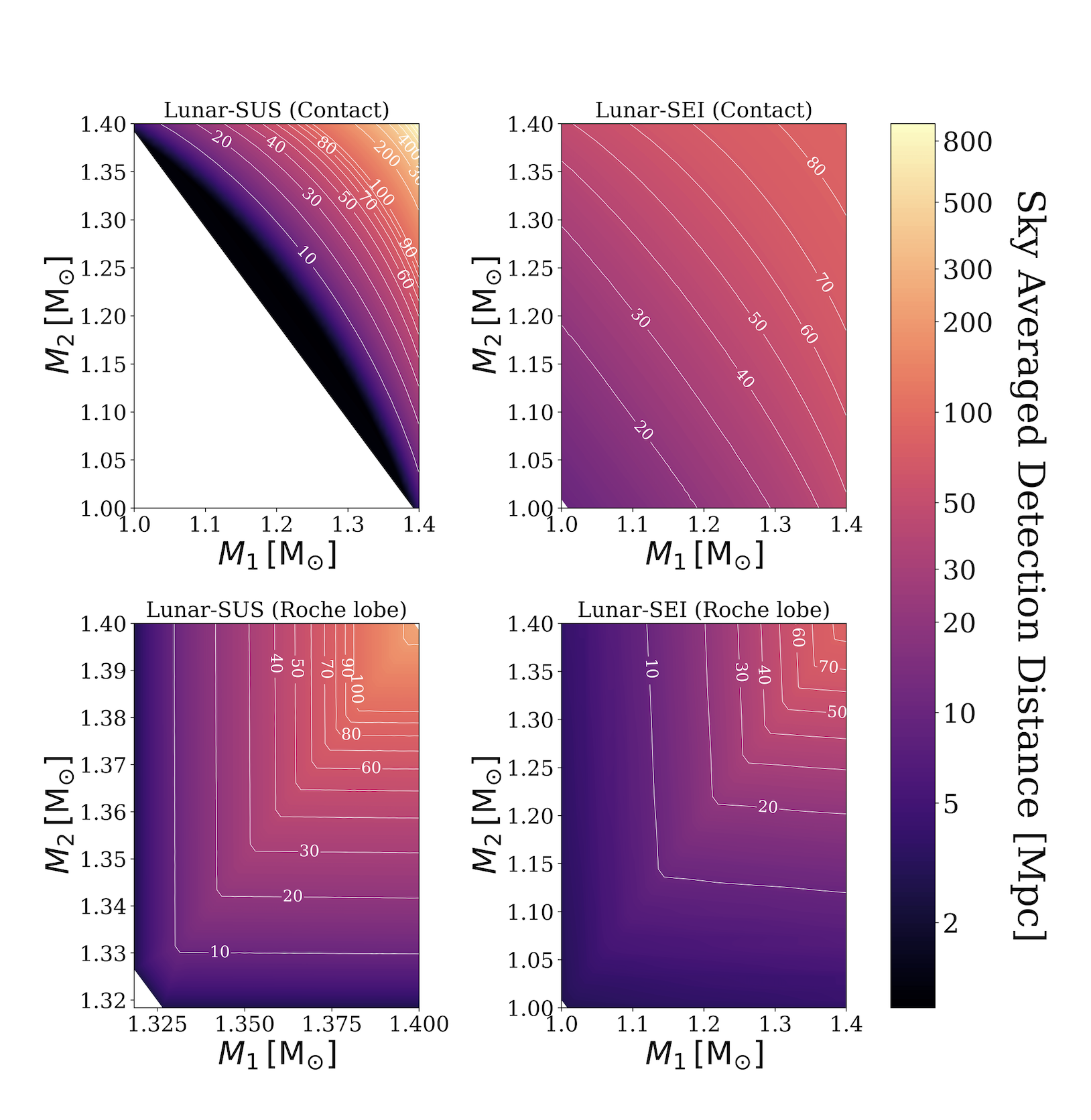}
\caption{Sky-averaged detection distances for all the WD mass combinations for \texttt{Lunar-SUS} and \texttt{Lunar-SEI} detectors, considering both contact (top panels) and Roche lobe (bottom panels) merger frequency definitions. The white contours indicate detection distances in Mpc, spanning from 1 to 800 Mpc as shown by the colorbar to the right. }  
\label{fig:HRDistAllMassComb}
\end{figure*}

Figure ~\ref{fig:HRDistBothCases} shows the sky-averaged detection distance for three different WD mass combinations. We assumed detection threshold of $\rho_{th} = 8$ (Equation~\ref{eq:HD}). We adopt an SNR threshold of 8 for detection, consistent with the standard used in gravitational-wave literature \citep[e.g.][]{Martynov(2016)PhRvD2016PhRvD..93k2004M}. Results are shown for both \texttt{Lunar-SUS} and \texttt{Lunar-SEI} detectors, comparing the contact case (solid bars) and Roche lobe overflow case (hatched bars) scenarios. LISA's detection range (orange) remains nearly constant at $\sim 0.4$ Mpc across all mass combinations. Figure~\ref{fig:HRDistAllMassComb} shows the sky-averaged detection distance for all the simulated WD mass combinations for both \texttt{Lunar-SUS} and \texttt{Lunar-SEI} detectors considering both contact and Roche lobe merger frequency definitions. We see that for the contact case and the most massive systems ($1.4~M_\odot+1.4~M_\odot$), \texttt{Lunar-SUS} has a detection distance 10 times that of \texttt{Lunar-SEI}. As the total mass of the system decreases, the detection distances for both detectors become comparable, with \texttt{Lunar-SEI} performing better than \texttt{Lunar-SUS} for less massive systems.

\subsection{Early Warning}

\begin{figure*}[t!]
\centering
\includegraphics[width=0.95\textwidth]{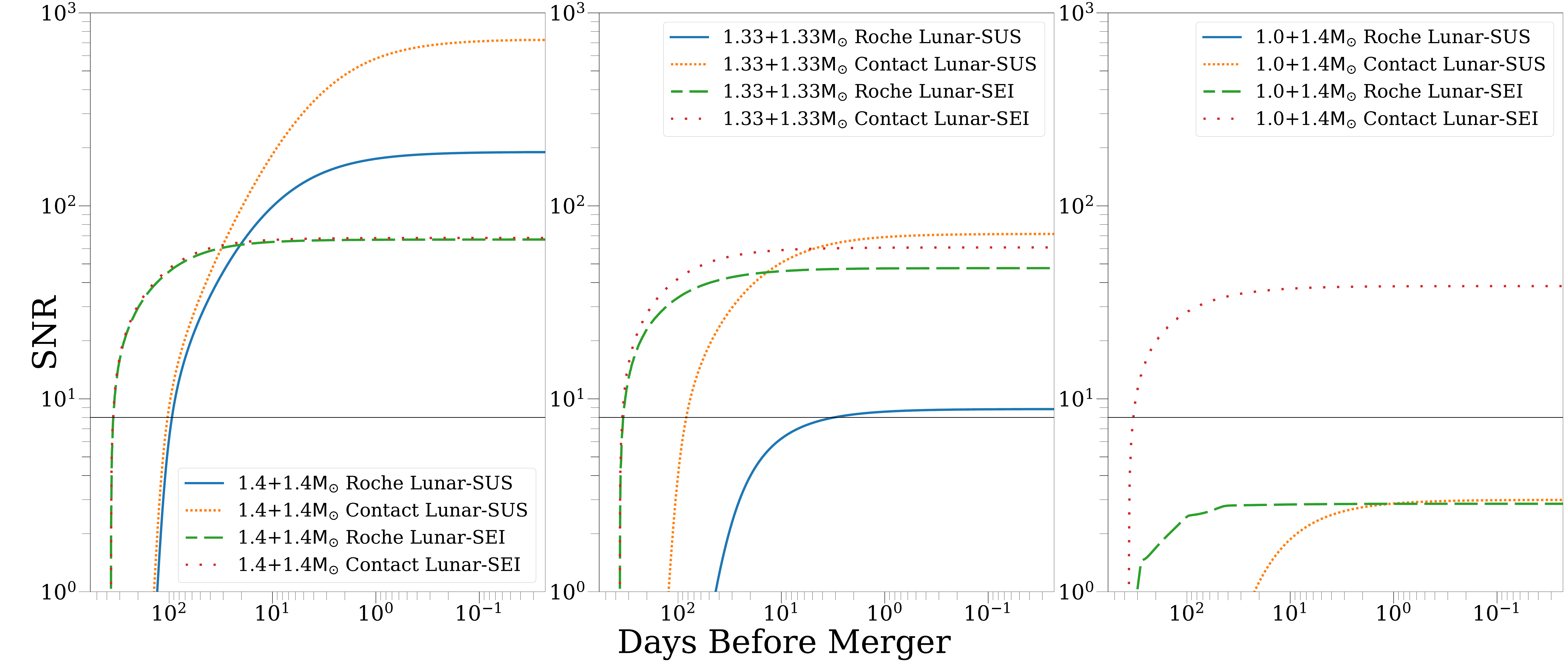}
\caption{Evolution of the SNR as a function of time before merger for three representative mass combinations: $1.4~M_\odot+1.4~M_\odot$(left panel), $1.33~M_\odot +1.33~M_\odot$ (middle panel), and $1~M_\odot +1.4~M_\odot$ (right panel), all at a reference distance of 10 Mpc. For each mass combination, we present results for both merger cases for \texttt{Lunar-SUS} (blue solid and orange dotted lines) and for \texttt{Lunar-SEI} (green dash-dotted and red dotted lines). The horizontal black line indicates our detection threshold of $\rho_{th}  =8$. We }
\label{fig:combinedAlert}
\end{figure*}

Figure~\ref{fig:combinedAlert} shows the evolution of the SNR as a function of time before merger for three representative mass combinations: $1.4~M_\odot+1.4~M_\odot$(left panel), $1.33~M_\odot+1.33~M_\odot$ (middle panel), and $1~M_\odot+1.4~M_\odot$ (right panel), all at a reference distance of 10 Mpc. 
To calculate the SNR at each point, we integrate Equation~\ref{eq:snr} setting $f_{min}$ as the frequency one year before merger and $f_{max}$ to be the frequency at $T$, where $T$ is the time before merger in days (X-axis). For each mass combination, we present results for both merger cases for \texttt{Lunar-SUS} (blue solid and orange dotted lines) and for \texttt{Lunar-SEI} (green dash-dotted and red dotted lines). The horizontal black line indicates our detection threshold of $\rho_{th}=8$. As expected, because of its low-frequency sensitivity, \texttt{Lunar-SEI} starts to observe the gravitational-wave signal from the double WD binary significantly earlier before the merger than \texttt{Lunar-SUS}.



\subsection{Parameter Estimation}



Table~\ref{tab:metrics} summarizes our parameter estimation analysis for three representative mass combinations. For each configuration, we report the 90\% credible levels for measurement uncertainties in sky localization ($\Delta\Omega$), luminosity distance ($\Delta D$) and the individual masses ($\Delta M_{1}$, $\Delta M_{2}$) for both \texttt{Lunar-SUS} and \texttt{Lunar-SEI} detectors. In this analysis, we consider $M_{1} \geq M_{2}$. For each binary system, we analyze both contact and Roche lobe merger scenarios. Results are derived using the Fisher matrix formalism implemented in GWFish. The gravitational-wave frequency evolution depends on two combinations of the masses $M_1$ and $M_2$ that can be well-measured and inverted to provide constraints on individual masses~\citep[See section II.C of ][for details on parameter estimation]{Cutler1994PhRvD..49.2658C}. 

For the $1~M_\odot+1.4~M_\odot$ combination, we see that \texttt{Lunar-SUS} does not detect the Roche lobe case while the \texttt{Lunar-SEI} sees some signal in the band but not enough to claim a detection. For the contact case, \texttt{Lunar-SEI} has an early-warning of almost a year before the merger with a sky-localization error of 0.0029 deg$^2$, hence helping in electromagnetic follow-up. It outperforms the \texttt{Lunar-SUS} detector which sees very little of the signal. We see that for the heaviest of the double WD binary systems, \texttt{Lunar-SEI} outperforms the \texttt{Lunar-SUS} detector by several orders of magnitude in term of sky-localization capabilities, which could be explained by the fact that the signal from this source spends a considerably longer time in the \texttt{Lunar-SEI} band than the \texttt{Lunar-SUS}. For the distance localization of the most massive case, both detectors have comparable errors for the Roche lobe case, whereas for the contact case \texttt{Lunar-SUS} gives twice as better estimate. 

The detectors are able to constrain the individual masses up to one part in a million for the systems we considered. We see that for equal mass systems, both detectors provide comparable constraints for the contact and Roche lobe cases, with \texttt{Lunar-SEI} performing an order of magnitude better than \texttt{Lunar-SUS}. 

Several physical uncertainties could potentially affect our parameter estimations. We assume circular orbits; this is justified as we expect eccentricity to be close to zero primarily due to tidal interactions \citep{Ruiter(2010)ApJ2010ApJ...717.1006R}. We use a simplified waveform model and calculate the expected gravitational wave strain under the assumption that the binary WD system can be represented by two point masses in a circular orbit. This approach is justified by \citet{vandenBroek(2012)MNRAS2012MNRAS.425L..24V}, who found that differences in the average strain amplitude due to deformations from Roche lobe filling or the existence of an accretion disk are typically at the level of one percent or less. For the parameter estimation, we use a Fisher matrix formalism that is only applicable for high SNR cases \citep{Vallisneri(2008)PhRvD2008PhRvD..77d2001V}.

\section{Discussion}
\label{sec:discussion}



\subsection{Expected Type Ia Rates}

One way to estimate the merger rate of double WDs is by assuming that all Type Ia supernovae result from double WD mergers. \citet{KinugawaDecihertztypeIa2022ApJ...938...52K}  used the Type Ia supernova rate in nearby galaxies to estimate a double WD merger rate within an 11 Mpc radius of $0.85 \, \text{yr}^{-1}$. Similar numbers were calculated by \citet{LGWAWP}, who determined the rate of Type Ia supernovae per year. They used the observed supernova sample from the Zwicky Transient Facility (ZTF) \citep{Bellm(2019)PASP2019PASP..131a8002B} survey and the Asteroid Terrestrial-impact Last Alert System (ATLAS) survey \citep{Tonry(2018)PASP2018PASP..130f4505T} from 2018 to 2023 and found 4 Type Ia supernovae within 10 Mpc. Adjusting for the average coverage of the surveys, they estimate  $1 \pm 0.4$ supernovae per year within 10 Mpc \citep{LGWAWP}. Other volumetric rates of Type Ia supernovae come from \cite{LiSN2011} which estimate a rate of  $0.25 \pm 0.05 \times 10^{-4} \, \rm Mpc^{-3} \, \rm yr^{-1}$ (about 0.1 Type Ia supernova per year at 10 Mpc), this estimate is also consistent with more recent studies like  \citet{CappellaroSN2015A&A...584A..62C}. We use this as our fiducial value  in the rest of the paper and should be seen as best case scenario (see Table~\ref{tab:rates}), especially since this approach ignores the contribution of the SD progenitor scenario to the overall Type Ia supernovae rate.

However, it's important to note that the merger rate is strongly dependent on the masses and types of the WDs \citep[e.g.][]{Cheng2020HighMassWDFromMerger}. Carbon-Oxygen (CO) WDs, proposed to be the most common progenitors of Type Ia supernova, can be as heavy as $1.25 M_{\odot}$ \citep{Hurley(2000)MNRAS2000MNRAS.315..543H}. We assume that the Type Ia rate corresponds to the  total rate of all CO+CO mergers and take it as a reference point. ONe WDs have masses in the range from $1.08 M_\odot$ to $M_{ch}$, and these are thought to be more probable progenitors to AIC, but \cite{Marquardt(2015)A&A2015A&A...580A.118M} showed that a significant fraction ($3-10\%$) of potential Type Ia progenitor systems should contain an ONe WD. To calculate an expected rate of Type Ia SN from massive WD binaries ($M_1,M_2> 1$) we use the relative rates for different WD types as reported in \citet{Marquardt(2015)A&A2015A&A...580A.118M} using the models of \citet{Ruiter(2014)MNRAS2014MNRAS.440L.101R}. \citet{Marquardt(2015)A&A2015A&A...580A.118M} calculates a relative rate of 0.27 for two CO+CO with primary mass (more massive WD) $> 0.9 M_\odot$. This could mean that instead of expecting one detection per year, we could expect one in 4 years or less from nearby events (10 Mpc). This is an overly optimistic scenario as there are fewer CO+CO where both the primary and the secondary are more massive than $1 M_\odot$. We can estimate a more realistic rate by considering only the rate of mergers of ONe+ONe WDs. \citet{Marquardt(2015)A&A2015A&A...580A.118M} calculates a relative rate of 0.04 for systems of ONe+X mergers (ONe+ONe and ONe+CO WDs combined), and state that about $14\%$ of those are ONe+ONe. This gives a relative rate of about 0.006, implying that to detect one event per year we need to reach horizon distances of $\sim 100$ Mpc. So to detect a substantial number of these high-mass mergers, \texttt{Lunar-SUS} and \texttt{Lunar-SEI} must achieve a sufficiently large detection distance. Extending the reach of these observatories is crucial for capturing a representative sample of double WD mergers across a wide range of masses. This will enable a more comprehensive understanding of the population and evolution of these systems, as well as the diverse outcomes of their mergers, including Type Ia supernovae.

\begin{deluxetable*}{lcccccccc}[t!]
\tablecaption{Detection Metrics for Lunar-Based gravitational-wave Observatories\label{tab:metrics}}
\tablewidth{0pt}
\tabletypesize{\footnotesize}
\tablehead{
\colhead{Masses} & 
\colhead{Detector} &
\colhead{Case} &
\colhead{SNR} & 
\colhead{$\Delta\Omega$} &
\colhead{$\Delta D$} &
\colhead{$\Delta M_{1}$} &
\colhead{$\Delta M_{2}$} &
\colhead{Early Warning} \\
\colhead{($M_\odot$)} &
\colhead{} &
\colhead{} &
\colhead{@10Mpc} & 
\colhead{(deg$^2$)} &
\colhead{(Mpc)} &
\colhead{($M_\odot$)} &
\colhead{($M_\odot$)} &
\colhead{(days)}
}
\decimals
\startdata
1 + 1.4 & \texttt{Lunar-SUS} & Contact & 3 & $2375$ & 10.4 & $3\times10^{-4}$ & $2\times10^{-4}$ & 0 \\
          &      & Roche   & -- & -- & -- & -- & -- & -- \\
          & \texttt{Lunar-SEI} & Contact & 38 & $2.9\times10^{-3}$ & $ 0.2$ & $3.9\times 10^{-6}$ & $ 2.2\times10^{-6}$ & 330 \\
          &      & Roche   & 3 & 1.8 & $ 2.9$ & $1.1\times10^{-4}$ & $ 5.3\times10^{-5}$ & 0 \\
\hline
1.33 + 1.33 & \texttt{Lunar-SUS} & Contact & 68 & 2.2 & $ 0.35$ & $ 8.8\times10^{-6}$ & $ 8.8\times10^{-6}$ & 82 \\
          &      & Roche   & 8.8 & 394 & $ 4$ & $5.3\times10^{-5}$ & $5.3\times10^{-5}$ & 3 \\
          & \texttt{Lunar-SEI} & Contact & 61 & $4.9\times10^{-4}$ & $ 0.12$ & $2.6\times10^{-7}$ & $2.6\times10^{-7}$ & 346 \\
          &      & Roche   & 47 & $1.8\times10^{-3}$ & $ 0.16$ & $1.6\times10^{-6}$ & $1.6\times10^{-6}$ & 338 \\
\hline
1.4 + 1.4 & \texttt{Lunar-SUS} & Contact & 750 & 0.03 & $ 0.04$ & $2.9\times10^{-7}$ & $2.9\times10^{-7}$ & 104 \\
          &      & Roche   & 206 & 0.28 & $ 0.12$ & $2.5\times10^{-6}$ & $2.5\times10^{-6}$ &  95 \\
          & \texttt{Lunar-SEI} & Contact & 68 &  $1.8\times10^{-4}$ & $ 0.11$ & $6.4\times10^{-8}$ & $6.4\times10^{-8}$ & 347 \\
          &      & Roche   & 67 & $2.8\times10^{-4}$ & 0.11 & $1.2\times10^{-7}$ & $1.2\times10^{-7}$ & 346 \\
\enddata
\tablecomments{SNR values are calculated for sources at 10 Mpc using Equation~\ref{eq:snr}. Error in sky localization ($\Delta\Omega$), luminosity distance ($\Delta D$) and masses ($\Delta M_{1}$, $\Delta M_{2}$) quoted at $90\%$ credible levels are calculated using the Fisher-matrix formalism in GWFish. Early warning time indicates how many days before the merger the source becomes detectable with an $\rho_{th}=8$. The contact case assumes merger at the point of physical contact between the WDs ($a = R_1 +R_2$), while the Roche lobe case considers merger at the point where the less massive WD fills the Roche lobe, see Equation~\ref{eq:gwroche}.}
\end{deluxetable*}

\subsection{WD mergers in GCs}

The contribution of GCs to the double WD merger rate is expected to be relatively small, but not entirely negligible, especially for future gravitational-wave observatories with increased sensitivity. \citet{Kremer2021WhiteDwarfGCs} used simulations to estimate the double WD merger rate in GCs, finding that it depends on the fraction of core-collapse clusters. In their simulation $90\%$ of mergers had a total mass greater than $M_{ch}$ and they considered WD masses from $0.2$ to $1.4 M_\odot$ \citep[see][Figure 7]{Kremer2021WhiteDwarfGCs}. They derived an overall WD merger rate of:

\begin{equation}  
 f \times 50 \text{ Gpc}^{-3} \text{ yr}^{-1} 
\label{eq:gcsrate}
\end{equation}

 where $f$ represents the fraction of core-collapse clusters. Assuming a typical core-collapse fraction of $20\%$ for Milky Way GCs \citep{Harris(1996)AJ1996AJ....112.1487H}, the estimated merger rate within 10 Mpc distance from Equation~\ref{eq:gcsrate} and taking $f=0.2$ is $\sim 10^{-5}\, \rm yr^{-1}$. This is taken as the fiducial value, but should be seen as the most optimistic rate that includes WDs of all masses. From \citet{Kremer2021WhiteDwarfGCs} simulations, roughly $44\%$ of mergers/collisions involve two CO WDs, $13\%$ involve two ONe WDs, and $38\%$ involve a CO+ONe WD pair. Therefore, when considering only mergers of the most massive WDs (ONe+ONe), the rate should be reduced to $13\%$ of the total merger rate. While this contribution is minor compared to the overall merger rate, it becomes more significant when considering the potential reach of future lunar-based observatories that could reach $D_{avg} > 300$ Mpc and see up to an event per year from a GC, see Figure~\ref{fig:HRDistAllMassComb}. This increased detection rate would provide valuable insights into the population and evolution of double WDs systems in GCs, as well as the role of dense stellar environments in the formation and merging of these systems.



\subsection{Accretion-Induced Collapse}

Another approach to estimating the expected rate of massive double WD mergers is to consider the predicted rate of AIC events. The expected progenitors of AIC events are ONe WD and their masses can range from $1.08 M_\odot$ to $M_{ch}$ \citep[see][]{Hurley(2000)MNRAS2000MNRAS.315..543H}, basically all the mass range considered in this study. The detection of an AIC event resulting from a double WD merger by a lunar-based gravitational-wave observatory would provide the first direct confirmation of this process as a viable mechanism for NS formation. 

The estimated rates of AIC events are highly sensitive to the parameters of common envelope (CE) ejection, which are currently not well-constrained and CE plays a critical role in binary stellar evolution \citep{1993PASP..105.1373I}. Using binary population synthesis models, \citet{Liu(2020)AICRates} estimated rates of AIC events from double WD mergers in the Galaxy to be in the range of $1.4{-}8.9\times 10^{\rm -3}\, \rm yr^{\rm -1}$. Lower estimates of $\lesssim 10 ^{-4}\, \rm yr^{\rm -1}$ come from earlier population synthesis studies \citep{Yungelson1998ApJ...497..168Y,Ruiter2019AICMNRAS.484..698R}. Constraints based on the amount of highly neutron-rich material ejected per AIC also suggest rates of $\lesssim 10 ^{-4}\, \rm yr^{\rm -1}$ \citep{Hartmann1985ApJ...297..837H, Fryer1999ApJ...516..892F, Metzger2009MNRAS.396.1659M}. Assuming that the AIC rate is proportional to the blue stellar luminosity \citep{Phinney1991Scaling}, an optimistic galactic rate of $10^{-3} \, \rm yr^{\rm -1}$ corresponds to a volumetric rate of $10^{-5} \, \rm Mpc^{-3}\, \rm yr^{-1}$. So at a distance of 30 Mpc we estimate $\cal O(\rm 1)$ events per year and $\cal O(\rm1000)$ events per year at a distance of 300 Mpc. \citet{Liu(2020)AICRates} studied the distribution of the total mass of merging double WDs that may result in an AIC. For the upper mass range, the most massive WD pairs ($1.4M_\odot + 1.4 M_\odot$), which have the largest detection distances, they calculated a relative frequency of 0.01. This means that, at 300 Mpc instead of thousands of event, we can expect to detect $\cal O(\rm 10)$  mergers of the most massive WDs. However, it's important to reiterate that these estimates strongly depend on the specific CE ejection models used in the calculations, and are mass dependent with mergers of more massive WDs being less frequent. This is also true for theoretical models of DWD mergers that could give rise to Type Ia SN \citep[e.g.][]{Toonen2012A&A...546A..70T,Maoz2018Merger}. 


Accurate determination of these rates is essential for understanding the contribution of AIC to the population of NSs and could potentially place constraints on CE models. Even if the number of Galactic NS population formed via AICs is expected to be $\le 0.1\%$ of the total \citep{Fryer1999WhatcanAICExplain}, this number could be significancy greater in dense stellar environments like GCs \citep{Ivanova(2008)MNRAS2008MNRAS.386..553I}.

Moreover, AIC events are proposed to play a role in the synthesis of heavy elements through r-process nucleosynthesis \citep{Fryer1999WhatcanAICExplain}. Therefore, constraining AIC rates not only informs models of compact object formation but also has substantial implications for our understanding of chemical evolution in galaxies.

\begin{deluxetable*}{llcccll}[t!]
\tablecaption{Expected Detection Rates for AIC and Type Ia SN Events\label{tab:rates}}
\tablewidth{0pt}
\tabletypesize{\footnotesize}
\tablehead{
\colhead{Event Type} &
\colhead{Population} & 
\colhead{Local Rate} & 
\colhead{Reference} &
\colhead{Masses} &
\colhead{\texttt{Lunar-SUS} Detections} & 
\colhead{\texttt{Lunar-SEI} Detections} \\
\colhead{} &
\colhead{} & 
\colhead{(Mpc$^{-3}$ yr$^{-1}$)} &
\colhead{} &
\colhead{($M_\odot$)} &
\colhead{(per yr)} &
\colhead{(per yr)}
}
\decimals
\startdata
Type Ia SN & Field & $0.25 \times 10^{-4}$ & \citet{LiSN2011} & $1+1.4$ & 0.006 [0] & 12 [0.005] \\
   &       &                               &                    & $1.33+1.33$ & 64 [0.1] & 45 [21] \\
   &       &                               &                    & $1.4+1.4$ & $9\times 10^{4}$ [1785] & 65 [62] \\
   & GCs & $1\times10^{-8}$ & \citet{Kremer2021WhiteDwarfGCs} & $1+1.4$ & $2\times 10^{-6}$ [0] & 0.005 [$2\times10^{-6}$] \\
   &       &                     &                        & $1.33+1.33$ & 0.025 [$4\times 10^{-5}$] & 0.018 [0.008] \\
   &       &                               &                    & $1.4+1.4$ & 35 [0.7] & 0.026 [0.024] \\
\hline
AIC & Field &$1\times 10^{-5} $ & \citet{Liu(2020)AICRates}  & $1+1.4$ & 0.002 [0] & 5 [0.002] \\
   &       &                               &                    & $1.33+1.33$ & 26 [0.04] & 18 [9] \\
      &       &                               &                    & $1.4+1.4$ & $4\times 10^{4}$ [714] & 26 [25] \\
   & GCs & $4.5\times10^{-8}$ & \cite{KremerFRBGCs2021ApJ...917L..11K} & $1+1.4$ & $1\times 10^{-5}$ [0] & 0.02 [$9\times10^{-6}$] \\
      &       &                               &                    & $1.33+1.33$ & 0.11 [$2\times 10^{-4}$] & 0.08 [0.038] \\
   &       &                     &                        & $1.4+1.4$ & 156 [3.2] & 0.12 [0.11] \\
\hline
\enddata
\tablecomments{Expected detections per year using the sky-averaged detection distances from Figure~\ref{fig:HRDistBothCases}. Numbers outside [inside] brackets correspond to Contact [Roche lobe] merger cases. }
\end{deluxetable*}

\subsection{AIC rates in GCs}

\citet{KremerFRBGCs2021ApJ...917L..11K,Kremer2023ApJ...944....6K} studied the dynamical formation scenarios for objects in old GCs that may plausibly power Fast radio bursts (FRBs) via an AIC event. Out of the $70\%$ of the WD mergers simulated for an M87 GC system, \citet{Kremer2023ApJ...944....6K} estimated that a fraction of order unity of these mergers may indeed lead to FRB sources. They estimated a super-Chandrasekhar WD + WD merger rate through tidal capture of up to roughly $7 \times 10^{-8} \, \rm yr^{-1}$ per typical core-collapsed GC and a volumetric rate of roughly $45 \,  \rm Gpc^{-3} \, \rm yr^{-1}$ in the local universe. For non-core collapsed this number could be 10 to 100 times smaller \citep[see e.g.][]{Kremer2021WhiteDwarfGCs} due to low central densities. 

We summarize the different rates (both for Type Ia supernova and AIC events) in Table~\ref{tab:rates} and estimate the expected number of events per year for each detector and merger case, assuming the most optimistic scenarios. For Type Ia SN, these are averaged volumetric rates that assume a relation between double WD mergers and Type Ia SN events. The first association of a gravitational-wave from a double WD merger would confirm this connection. For AIC rates, these are theoretical models with significant assumptions. Detecting the first mergers of double WDs would be similar to the first observations of double NS mergers \citep{Abbott2017PhRvL.119p1101A} with LIGO, where the predicted rates were highly different from the actual observations. To date, only one such event has been confirmed with an additional candidate \citep{Abbott2020ApJ...892L...3A}.


If a merger event is detected by \texttt{Lunar-SEI} or \texttt{Lunar-SUS} and results in an AIC there are several predictions that this event could be a multi-messenger event. 
%
%
AIC events in both the SD and DD models are expected to be radio-bright phenomena \citep[e.g.,][]{Piro2013ApJ...762L..17P,Moriya(2016)RadioAIC}. In the DD scenario, an AIC event can produce supermassive and rapidly rotating NSs with masses exceeding $2.2~M_\odot$ \citep[e.g.][]{Metzger2015Magnetar}. These NSs are potential progenitors of FRBs, which could be observable out to redshifts of $z > 0.7$ \citep{Falcke(2014)FRBsMassNSs,Moriya(2016)RadioAIC}. The discovery of repeating FRB \citep{Spitler(2016)RepeatFRB,Scholz(2016)FRBrepeattwo} means that FRB models with cataclysmic phenomena cannot account for all FRBs.       \citet{Chen(2024)FRBFormationRate} found a local rate of $1.13 \times  10^{4} \text{ Gpc}^{-3} \text{ yr}^{-1}$. If a FRB is detected following a merger observed by \texttt{Lunar-SEI} or \texttt{Lunar-SUS}, it would provide strong confirmation for this so-called ``blitzar" model \citep{Falcke(2014)FRBsMassNSs,Moriya(2016)RadioAIC,2020ApJ...893....9Z}. Such a model may only account to a small fraction of FRBs \citep[events in the high-luminosity end][]{Luo2020MNRAS.494..665L}.

A recent radio transient in M81 reported by \cite{Anderson(2019)RadioTransient}  has been attributed possibly to an AIC event of a WD by \cite{Moriya(2019)AICRadio} (considering both SD or DD channel). Additionally, FRB 121102 \citep{Scholz(2016)FRB} has been suggested to be associated with an NS formed via an AIC event possible from a SD or DD channel \citep{Margalit(2019)FRBfromAIC}.



AIC events are also potential X-ray and optical transients. \citet{Yu(2015)RapidEvolve} and \citet{Yu(2019)XrayTransientAIC} studied the hard and soft X-ray emission from an AIC event. In optical, a certain type of optical transients called ``Fast blue optical transients"\citep[FBOTs;][]{Drout2014ApJ...794...23D,Arcavi2016ApJ...819...35A,2018MNRAS.481..894P,Ho(2023)FBRT} could be explained by the AIC scenario. Arguably the most studied FBOT is the multi-wavelength event AT2018cow \citep{PrenticeCow2018ApJ...865L...3P,RiveraCown2018MNRAS.480L.146R,HoCow2019ApJ...871...73H,KuinCow2019MNRAS.487.2505K,MarguttiCow2019ApJ...872...18M,Perley2019MNRAS.484.1031P,NayanaCow2021ApJ...912L...9N}. For this event,  \citet{Lyutikov(2019)CowasAIC} argued that it could have been the result of a merger of an ONe WD (massive WD) with another WD. Additionally, \citet{McBrien(2019)AICtransient} identified SN2018kzr as a rapidly declining transient resulting from the destruction of a WD via an AIC event. 




Even at higher energies, we could expect to detect AIC events. Several authors have discussed the possibility that an AIC event could produce short Gamma-Ray Bursts \citep[GRBs;][]{DuncanAIC1992ApJ...392L...9D,VietriAIC1999ApJ...527L..43V,DessartAIC2007ApJ...669..585D,MetzgerAIC2009MNRAS.396.1659M,KingAIC2001MNRAS.320L..45K,LevanGammamagnet2006MNRAS.368L...1L,LeeAIC2007NJPh....9...17L,Piro2013ApJ...762L..17P,Lyutikov(2017)GammaRay}. More recently, \citet{Cheong2025ApJ...978L..38C} presented relativistic simulations of an AIC originating from double WD mergers and showed that they can generate relativistic jets and neutron-rich outflows with properties consistent with long GRB accompanied by kilonovae, such as GRB 211211A and GRB 230307A.

Overall, the detection of AIC events from DD mergers offers significant opportunities for multi-wavelength and multi-messenger astronomy. The ability to observe these events across a broad range of wavelengths—from radio to gamma rays—as early as a couple of weeks before the merger, depending on the system's mass and distance, can greatly enhance our understanding of these phenomena. This early detection capability allows for coordinated observations with various electromagnetic observatories, facilitating a comprehensive study of the merger process and the subsequent formation of NS or other compact objects.

\subsection{Multi-band gravitational-wave Astronomy}

Some double WD mergers may result in the formation of a rapidly rotating NS through AIC than then could then become sources of gravitational waves at the higher frequencies detectable by LIGO and future ground-based observatories \citep{Abdikamalov2010AICGWLIGO,LongoMicchi(2023)MNRAS}. More probable in GCs \citep{Ivanova(2008)MNRAS2008MNRAS.386..553I}. \citet{LongoMicchi(2023)MNRAS} estimate that  fast rotating AICs could be detectable up to a distance of 8 Mpc with third-generation GW observatories, and up to 1 Mpc with LIGO. \cite{Kremer2023ApJ...944....6K} found that $70\%$ of the WD mergers in their simulation of an M87-like GC system sample have properties consistent with a NS formation outcome, and estimate a WD merger rate of $260 \times 10^{-7} \rm yr^{-1}$. This means that the detection of binary WD mergers by lunar-based gravitational-wave observatories could have important synergies with ground-based detectors such as LIGO, and future ground-based gravitational-wave observatories.


\subsection{WDs and NS/BH mergers}

In addition to double WD mergers, lunar-based gravitational-wave observatories like GLOC, LGWA, and LILA could also detect mergers involving WDs and other compact objects, such as NS and BHs \citep{LGWAWP,LGWA(2021)}. For example, \citet{YamamotoWDsPlusBHprimal2024PhRvD.109j3514Y} investigated the potential for detecting primordial BH and WDs mergers with sub-Hertz detectors like DECIGO and found an expected detection rate of $\cal O(\rm 10^{-6})$ for primordial BH of $M_{PBH} \sim 0.1 M_\odot$. WD-BH mergers are also potential multi-messenger sources as some mergers could result in electromagnetic transients. \citet{Fryer(1999)ApJ1999ApJ...520..650F} estimated a rate as high as $10^{-6} \rm yr^{-1}$ per galaxy, and \citet{Dong(2018)MNRAS2018MNRAS.475L.101D} estimated a similar rate of $2\times 10^{-6} \rm yr^{-1}$ in the galactic disk for BH masses around $5-7 M_\odot$. \citet{LloydWDBH2024MNRAS.535.2800L} suggest that these mergers may constitute a significant subpopulation of low-redshift ($z <2$) long GRBs. The detection of a WD-BH in both gravitational waves and electromagnetic emission would provide a unique opportunity to study the physics of these extreme events and their role in the production of long GRBs.

Another possibility is the merger of a WD with an NS. These systems, like WD-BH mergers, are expected to be less common than double WD binaries but could still be detectable by lunar-based gravitational-wave observatories with sufficiently large horizon distances \citep{Bobrick(2017)MNRAS2017MNRAS.467.3556B,Toonen(2018)A&A2018A&A...619A..53T,Sberna2021ApJ...908....1S}. WD-NS mergers are also expected to produce electromagnetic counterparts \citep[e.g.][]{Bobrick(2022)MNRAS}, although the specific signatures may differ from those of double WDs or WD-BH mergers. The same rough SNR calculations used to estimate the early alert and detection distances can be applied to WD-NS/BH systems.



\section{Conclusion}
\label{sec:conclusion}




In this work, we explored the capabilities of future lunar-based gravitational-wave experiments in observing massive ($M_1, M_2 > 1~M_\odot$) double WD binary mergers. We use GLOC as a proxy for suspended test-mass detectors and LGWA as a representative of seismometer-based detectors on the Moon. Considering both contact and Roche lobe overflow merger scenarios, we find detection horizons extending close to ${\sim 1}~\rm Gpc$ for the most massive systems. Our calculations predict that lunar detectors could observe dozens to thousands of massive WD mergers annually, with additional contributions from GCs. The masses of these WDs would be constrained with an unprecedented accuracy of one part in a million. These detectors will provide early warnings weeks to months before merger. The sky-localization of those sources would be at a few square arcminute resolution, enabling a new era of coordinated multi-messenger follow-up of electromagnetic transients—whether they evolve into Type Ia supernovae or AIC events.

\section*{Acknowledgments}
\begin{acknowledgments}
We thank the anonymous referee for comments that improved the quality of this article. MPM acknowledges support from the EMIT NSF grant (NSF NRT-2125764). KJ and ABY acknowledge support from the Scaling Grant by the Vanderbilt Office of the Vice-Provost for Research and Innovation. This research has made use of the Science Explorer (SciX), funded by NASA under Cooperative Agreement 80NSSC21M00561. 
MPM would like to thank Diana Moreno-Santillan, Francisco Castellanos and Jose ``Pepe" Orozco for their support and interesting discussions. 
\end{acknowledgments}

%

\vspace{5mm}


\software{LEGWORK \citep{LEGWORK_apjs}, GWFish \citep{GWFishPaper}
          }





\bibliography{sample631}{}
\bibliographystyle{aasjournal}



\end{document}